\documentclass[11pt,letterpaper]{article}
\usepackage[utf8]{inputenc}
\usepackage[T1]{fontenc}
\usepackage{lmodern}
\usepackage{amsmath,amssymb}
\usepackage{graphicx}
\usepackage[letterpaper,margin=1in]{geometry}
\usepackage{authblk}
\usepackage{hyperref}
\usepackage{caption}

\title{Modeling the Monty Hall Decision Problem with\\Reaction Kinetics\protect\footnotemark[2]}

\author[1]{Oliver Steinbock\protect\footnotemark[1]}
\author[1]{Wen Zhu\protect\footnotemark[3]}
\affil[1]{Department of Chemistry and Biochemistry,\protect\\
  Florida State University, Tallahassee, FL 32306-4390, USA}

\date{}

\begin{document}
\maketitle

% footnotes in the right order:
\footnotetext[1]{\href{mailto:osteinbock@fsu.edu}{osteinbock@fsu.edu}}
\footnotetext[3]{\href{mailto:wzhu@chem.fsu.edu}{wzhu@chem.fsu.edu}}
\footnotetext[2]{Electronic supplementary information (ESI) available: Movie\,S1 (always-switch animation, MP4, 7~MB) and Movie\,S2 (always-stay animation, MP4, 10~MB). See DOI:...}

\begin{abstract}
\noindent Using the Monty Hall probability problem as a model system, we ask whether simple chemical reaction mechanisms can implement optimal strategies for non‐trivial decision making. In this puzzle, a contestant chooses one of three doors (only one hides a prize), the host—knowing the content—opens another door revealing no prize, and finally the contestant must decide whether to stay with the original choice or switch to the remaining closed door. Assigning distinct molecular species to the player, initial choice, reveal step, and final decision, we encode the problem in mass-action kinetics. For pseudo-first-order conditions, tuning a single rate constant shifts the network continuously between ``always-stay'' (1/3 success) and ``always-switch'' (2/3 success) regimes. We derive closed-form, time-dependent expressions for the success kinetics, concluding with a brief discussion of proposed DNA strand-displacement implementations and kinetically hard-wired molecular decision-making.
\end{abstract}

\vspace{1\baselineskip}

\vspace{1\baselineskip}
\begin{figure}[ht]
  \centering
  \captionsetup{labelformat=empty}    % no “Figure 1:” prefix
  \includegraphics[width=8cm]{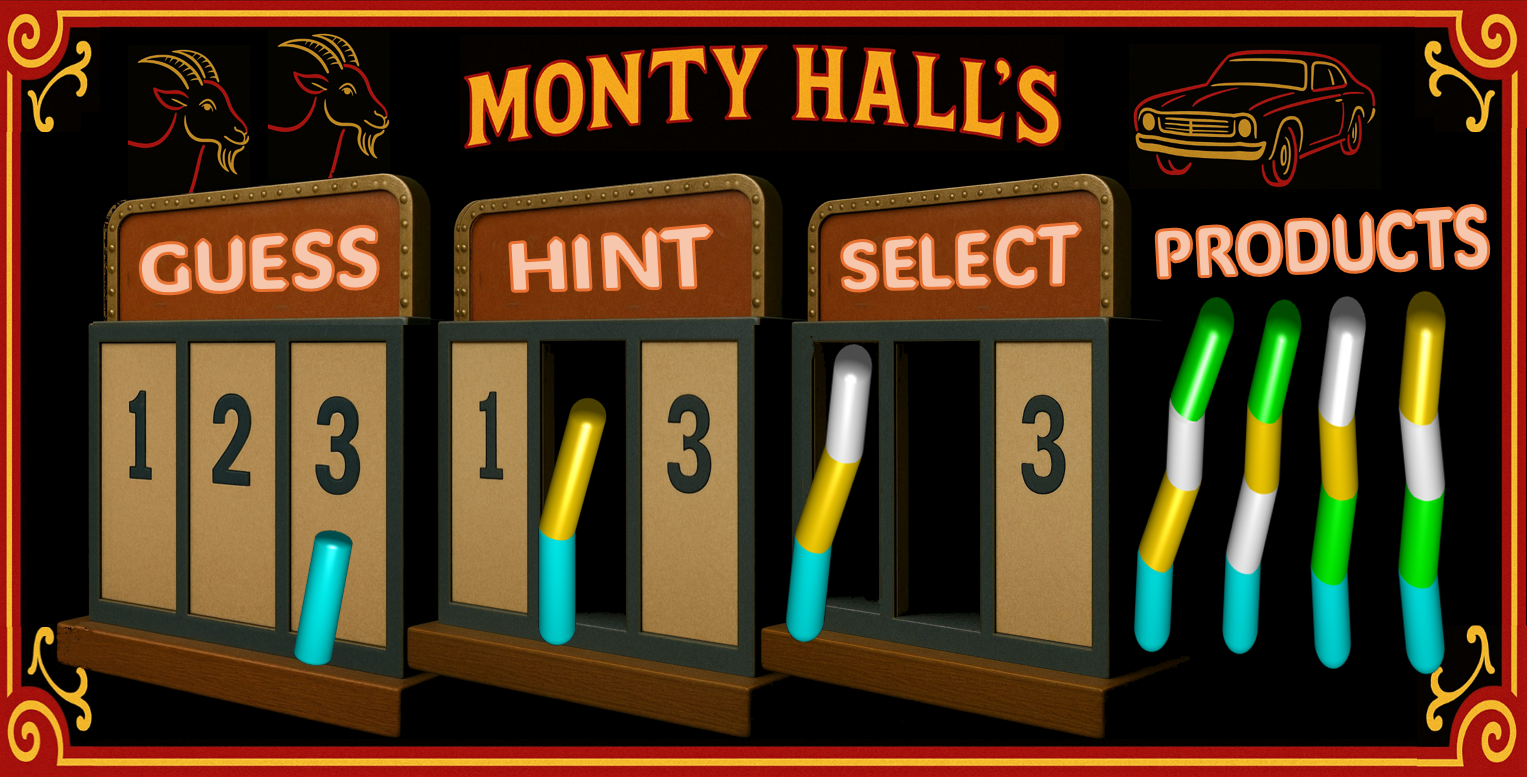}
  \caption{ToC Entry: A chemical reaction network mimics the Monty Hall problem, shifting decision outcomes by tuning kinetic rate constants.}
  \label{fig:ToC}
\end{figure}
\addtocounter{figure}{-1}  % next figure will be #1

\newpage

\section{Introduction}

Living systems continuously make decisions in response to environmental stimuli \cite{tyson2014}.
At the molecular scale, such decision-making processes are mediated by chemical reactions that regulate cellular responses. For example, stem cells select differentiation pathways based on external signals, and amoebae weigh chemoattractant versus chemorepellent signals when navigating complex environments \cite{costa2012,kay2008,friedl2003}. While much current research focuses on identifying molecular players within these regulatory networks, the deeper essence of decision-making lies in the underlying reaction mechanisms and their dynamics.

Chemical reactions can process information by utilizing nonlinearities in their rate laws. A classic example is the Hill kinetics, where sigmoidal response curves create switch-like behaviors: low substrate concentrations yield minimal output, but once a threshold is crossed, the output rapidly saturates \cite{delvecchio2016}. Such mechanisms effectively implement logical true/false decisions, enabling cells to filter noise and make robust choices between discrete fates.

Inspired by this capacity, numerous chemical systems have been designed to implement logic gates, finite-state machines, and neural network behaviors \cite{hjelmfelt1991,hjelmfelt1992a,hjelmfelt1992b}. Even relatively weak nonlinearities, such as second-order reaction terms, are sufficient to drive rich computational behaviors. Further complexity arises in reaction-diffusion systems \cite{zhu2024}, where chemical patterns can encode spatial decisions, including shortest-path selection \cite{steinbock1995} and midcell positioning in bacterial division \cite{beta2017}.

While most studies focus on thresholding behaviors and bistability, chemical systems may also be capable of implementing more subtle, probabilistic decision strategies. In this study, we investigate whether simple reaction networks can realize such strategies for counterintuitive problems in probability theory. As a model challenge, we focus on the Monty Hall problem \cite{selvin1975a}, a well-known puzzle where an optimal strategy emerges through non-obvious reasoning.

We show that the key elements of the Monty Hall game and its solution can be faithfully encoded in a set of linear, pseudo-first-order rate laws. By tuning a single rate constant, the reaction network continuously interpolates between ``always stay'' and ``always switch'' behaviors, achieving success concentrations of 1/3 and 2/3, respectively. This kinetic framework not only reproduces the characteristic outcomes of the Monty Hall problem but also offers a mechanistic route for experimental realization and pathway-encoded molecular strategies.

In this work, we use the term chemical reaction to refer broadly to abstract interaction rules between species governed by mass-action kinetics. Similarly, our use of reaction mechanism follows the convention in physical chemistry and systems chemistry, where a mechanism denotes a network of species and reaction steps characterized by rate laws, rather than a detailed molecular-level pathway involving orbitals, transition states, or bond rearrangements. This abstraction allows us to represent logic and strategy through kinetic flows and branching probabilities.

\section{The Monty Hall Problem}
The Monty Hall problem is a classic puzzle in probability and decision theory \cite{selvin1975a,whitmeyer2017,rosenhouse2019}. A contestant is presented with three closed doors: behind one is a prize (e.g., a car), and behind the other two are non-prizes (e.g., goats). The contestant selects one door. Next, the host who knows what is behind each door, opens one of the two remaining doors to reveal a goat. The contestant is then given the option to either stay with their original choice or switch to the other unopened door.

At first glance, one might conclude that retaining or switching yields equal odds \cite{krauss2003}; even the famed mathematician Paul Erdős reportedly refused to accept the advantage of switching until he was convinced by a Monte Carlo simulation \cite{wikipedia}. A Bayesian analysis, however, makes this advantage clear.  
When the contestant initially selects door~1, each door has a probability of \(1/3\).  
The two unchosen doors together therefore carry the remaining probability of \(2/3\).  
When the host opens one of those unchosen doors to reveal a goat, the sole remaining door inherits the full \(2/3\) probability.  
Thus, switching raises the win probability to \(2/3\), whereas staying leaves it at \(1/3\).  

For further illustration, consider a million doors. After the initial choice, the host opens 999{,}998 goat doors, leaving only the initial guess and one other door unopened.  Since the initial choice had only a \(10^{-6}\) probability of concealing the prize, the remaining unopened door must carry the residual probability of 1 minus \(10^{-6}\).  Thus, switching increases the chance of winning from \(10^{-6}\) to essentially 1.

\section{Reaction Mechanisms}

We first identify the different components of the Monty Hall problem with distinct chemical species.  The three doors are denoted as species \(D_j\) (\(j = 1,2,3\)), and the initial player as \(P\).  When a \(P\) molecule makes its initial choice \(D_j\), it reacts to form \(I_j\).  After the host reveals the wrong choice \(D_k\), \(I_j\) reacts further to \(I_{j,k}\).  Finally, the selection of the remaining unopened door \(D_\ell\) yields the ultimate product species \(I_{j,k,\ell}\).

\begin{figure}[ht]
  \centering
  \includegraphics[width=0.6\linewidth]{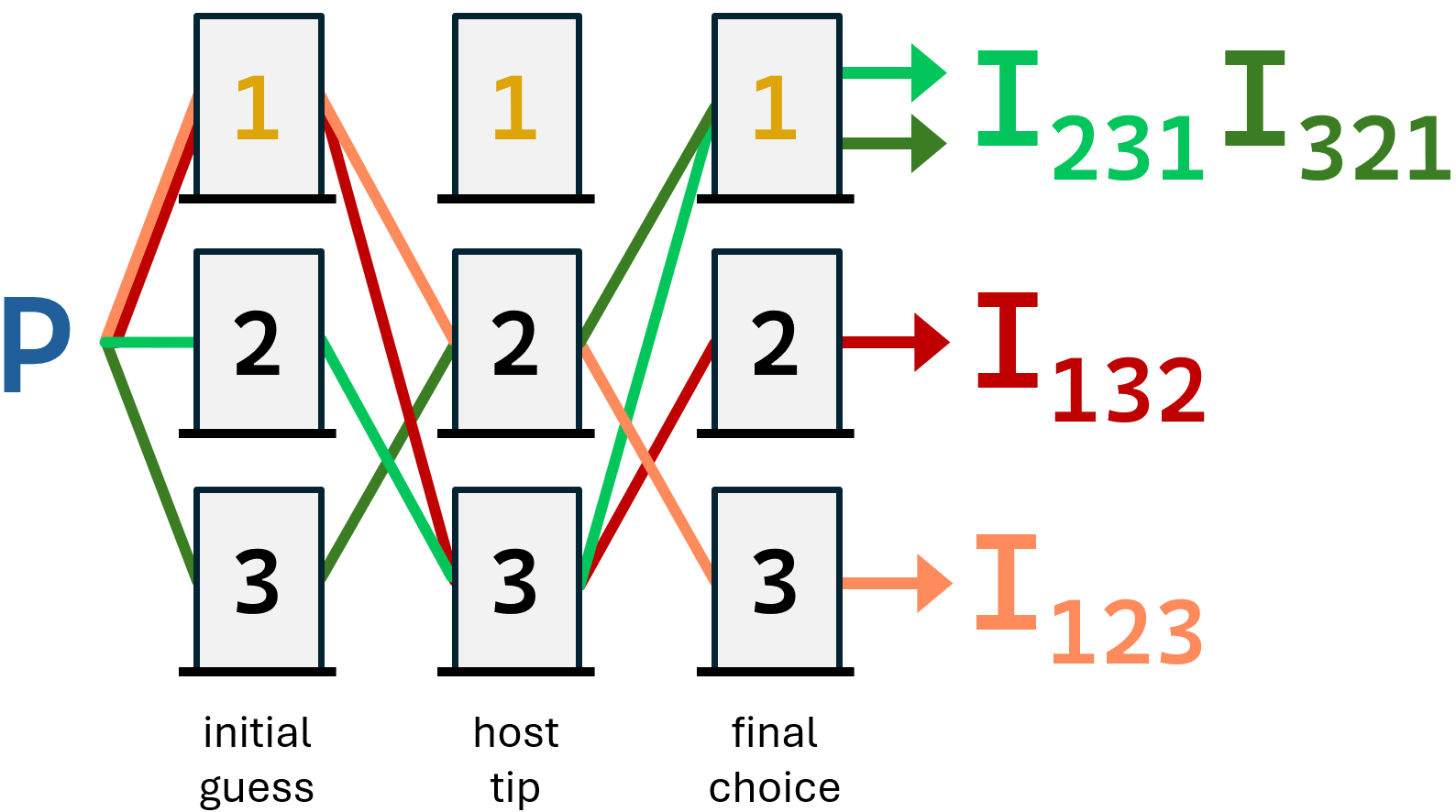}
\caption{Schematic of the four possible always‐switch pathways taken by the initial player species \(P\) through three sequential door interactions. In this example, the prize is behind door 1. Accordingly, only the green reaction pathways lead to winning products, while the red pathways result in losses.}
\label{fig:Schematic}
\end{figure}

In the following, we denote the winning door choice as \(g\) and, without loss of generality, assume \(g=1\).  This simplification is valid because the labeling of the doors is arbitrary; relabeling them does not affect the structure or outcome of the decision process.  Consequently, the final product species \(I_{j,k,\ell}\) represents wins if \(\ell=1\) and losses if \(\ell=2\) or \(3\).

The irreversible reaction steps of the always-switch mechanism are as follows:
\begin{align}
P + D_j &\xrightarrow{k_1} I_j 
  \quad\text{(initial guess)}\\
I_j + D_k &\xrightarrow{k_2} I_{j,k}, 
  \quad k\neq j,\;k\neq g
  \quad\text{(host presents goat)}\\
I_{j,k} + D_\ell &\xrightarrow{k_{\mathrm{sw}}} I_{j,k,\ell}, 
  \quad \ell\neq j,\;\ell\neq k
  \quad\text{(always switch)}
  \label{eq:swtch}
\end{align}

For \(g = 1\), this three-step mechanism consists of 11 reactions involving 15 species (see Figure~\ref{fig:Schematic} and Appendix A). Since the host never reveals the prize door \(D_1\), only four final product species are possible: \(I_{1,2,3}\), \(I_{1,3,2}\), \(I_{2,3,1}\), and \(I_{3,2,1}\), with the latter two being winning species. For identical rate constants \(k_1\), \(k_2\), and \(k_{\mathrm{sw}}\) and identical door concentrations, this mechanism is expected to yield the winning species at a concentration equal to two‐thirds of the initial concentration \([P]_0 = P_0\). As in the classic Monty Hall problem, however, the two‐thirds outcome is not intuitively evident. We note that if the initial door concentrations are unequal, the reaction network no longer reflects an unbiased Monty Hall game. For instance, if the prize door is in excess, the player tends to pick it first, inverting the usual advantage of switching. Thus, equal door concentrations are essential for faithfully modeling the decision symmetry.

While the above mechanism captures the always-switch strategy, alternative approaches can reflect different decision behaviors.  For instance, we can add a parallel ``stay'' reaction
\begin{equation}
\label{eq:stay}
I_{j,k} + D_j \xrightarrow{k_{\mathrm{st}}} I_{j,k,j}
  \quad\text{(stay)}
\end{equation}
which introduces four additional product species that correspond to a stay outcome 
\(
(I_{1,2,1},\;I_{1,3,1},\\I_{2,3,2},\;I_{3,2,3})\
\). 
By assigning different rate constants \(k_{\mathrm{sw}}\) and \(k_{\mathrm{st}}\) to the switch and stay reactions, the network captures a full spectrum of decision strategies through reaction kinetics. Two Monte Carlo animations, illustrating the kinetics of the always-switch and always-stay scenarios, are provided as ESI material.

\section{Rate Laws}
\label{sec:rate-laws}

We derive the rate laws of our Monty Hall mechanism under the assumption of bimolecular interaction events. Second-order rate laws then follow directly from mass-action kinetics, but the game rules impose additional constraints. Below, we outline key rate terms and their origins. The complete set of rate laws is given in Appendix B.

The player molecules \(P\) react with the three door species at a rate of
\(
r = k_{1}\,[P]\,[D_i] \
\)
\noindent to produce the three intermediates \(I_j\). When all door concentrations start equal, \(P\) is consumed three times faster than each individual \(I_j\) is formed. This scheme captures the simplest kinetic always-stay strategy; however, we will consider the more game-related mechanism that branches at stage-3 (see reaction~\ref{eq:stay}).

The consumption terms in the rate laws of the first-stage intermediates \(I_j\), which represent the host’s hint, are more contracted as \textit{i}) the winning door~1 is not involved and \textit{ii}) the door tip is always different from the player's initial choice. For example, \(I_2\) will have a consumption rate of 
\(
r = k_{2}\,[I_2]\,[D_3] \
\), 
while \(I_1\) follows 
\(
r = k_{2}\,[I_1]\,([D_2]+[D_3]) \
\).

The third-stage rate terms describe the ultimate decision. At this point, the relevant reactant species \(I_{j,k}\) had two door interactions and the always-switch rule allows only for a reaction with the door species \(D_\ell\) for which \(\ell\) is neither \(j\) nor \(k\). Accordingly, there is exactly one consumption term for each \(I_{j,k}\) and the corresponding rate is 
\(
r = k_{\mathrm{sw}} [D_{\ell}] [I_{j,k}] \
\).

As shown in Figure~\ref{fig:Schematic}, four reaction paths connect \(P\) to the final products, of which two win and two lose.  However, because the host never reveals the prize door \(D_g\), the single path starting from the correct choice \(P + D_g\) splits into two losing outcomes, while each of the two paths starting from a goat door funnels into the sole remaining prize door.  Mass-action kinetics thus directs two‐thirds of the total flux into winning products and one‐third into losers, reproducing the classic 2/3 success probability.

When both the switch step (reaction~(\ref{eq:swtch})) and the stay step (reaction~(\ref{eq:stay})) are active, each intermediate \(I_{j,k}\) is consumed by two competing channels. Accordingly, the mass-action rate law for \([I_{j,k}]\) has now two consumption terms, namely
\(
r = k_{\mathrm{sw}}\,[I_{j,k}]\,[D_\ell] + k_{\mathrm{st}}\,[I_{j,k}]\,[D_j].
\)
This formulation recasts the Monty Hall problem and its two canonical strategies as a deterministic chemical reaction network.

\begin{figure}[ht]
  \centering
  \includegraphics[width=0.5\textwidth]{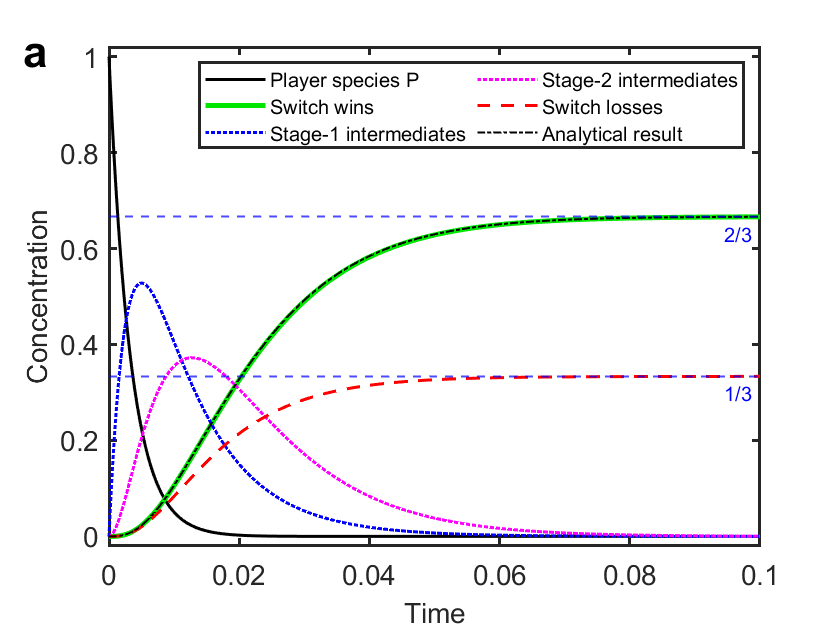}\par\medskip
  \includegraphics[width=0.5\textwidth]{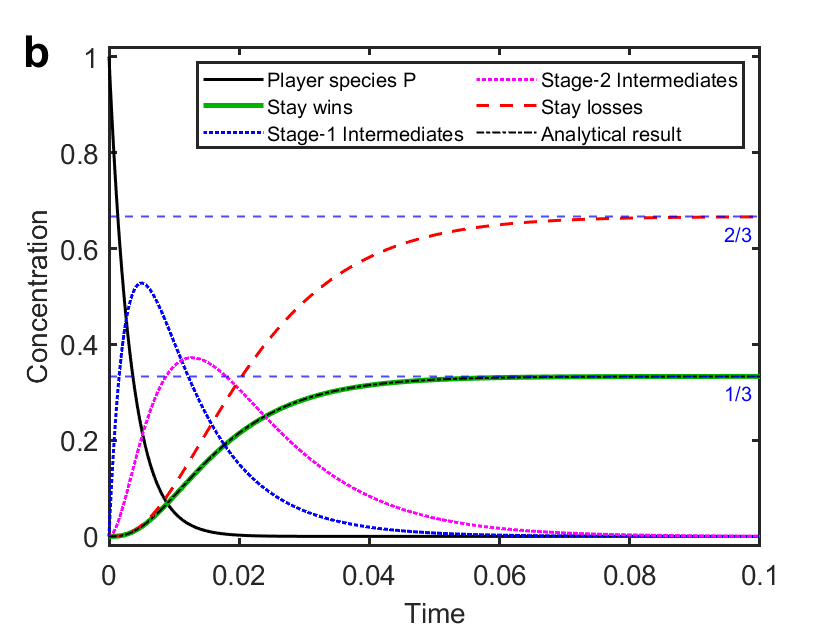}
  \caption{Numerical simulations of the (a) always-switch and (b) always-stay model. The total wins (green) and losses (red) approach the expected equilibrium values. The dotted blue and magenta curves show the time courses of the total concentrations of the three first-stage intermediates \(I_{j}\) and four second-stage intermediate species \(I_{j,k}\), respectively. The black dashed curves are the analytical results based on eq.~\eqref{eq:twoequal} (scaled in (b)). All rate constants \(k = 1\), except for \(k_{\mathrm{st}} = 0\) in (a) and \(k_{\mathrm{sw}} = 0\) in (b). Initial concentrations: \([P]_{0} = 1\) and \([D_{j}]_0 = 100\) for \(j=1,2,3\).}
  \label{fig:Kinetics}
\end{figure}

\section{Numerical Simulations}
\label{sec:numerical}
Figure~\ref{fig:Kinetics}a,b show the time evolution of the key species concentrations in the always‐stay and always‐switch mechanisms, respectively. Initial concentrations are set to 1 for \(P\), 100 for each \(D_j\), and zero for all other species. All rate constants equal 1. Time and concentration are treated as dimensionless quantities but can be assigned physical units if used consistently.

As expected, the concentration of \(P\) decays exponentially with an effective rate constant of \(3 k_1 [D]_0\) and a corresponding half-life of \(\ln(2)/300 \approx 0.0023\) time units. The intermediate species \([I_j]\) and \([I_{j,k}]\) rise and fall, ultimately supporting the formation of the final products. Most importantly, we observe that the always‐stay mechanism produces a final winner concentration of \(1/3\), while the always‐switch mechanism yields a value of \(2/3\). These final concentrations correspond exactly to the expected success probabilities of the Monty Hall problem.

\begin{figure}[ht]
  \centering
  \includegraphics[width=0.5\textwidth]{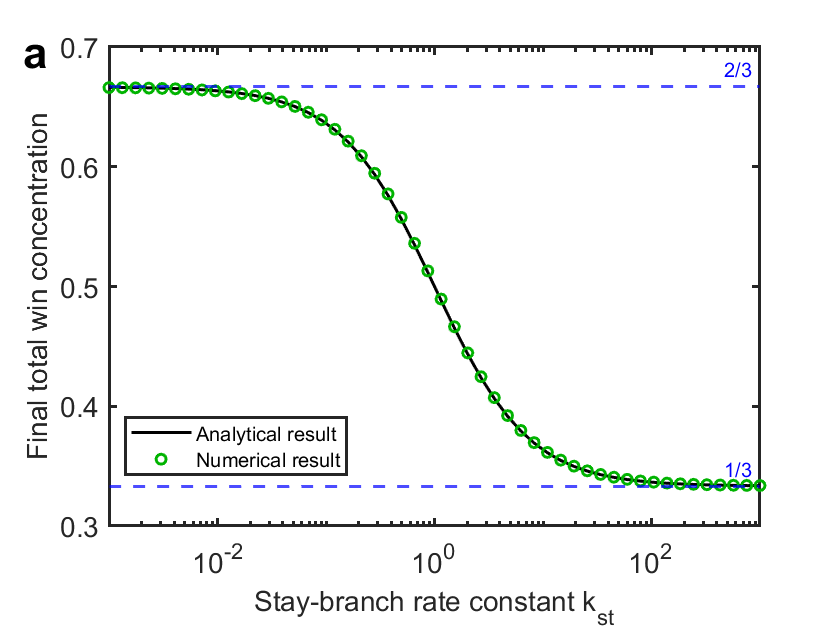}\par\medskip
  \includegraphics[width=0.5\textwidth]{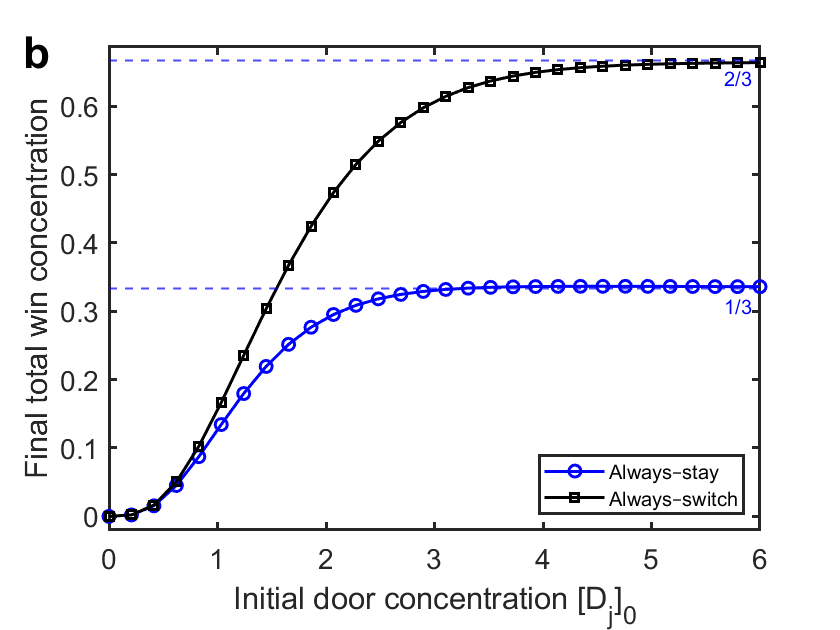}
\caption{(a) Equilibrium concentration of the winning species, $W_\infty$, as a function of the stay‐branch rate constant $k_{\mathrm{st}}$, showing numerical data (markers) and the analytical solution from eq.~\eqref{eq:Winf} (solid line). Parameters: $[P]_0=1$, $k_{\mathrm{sw}}=k_1=k_2=1$, and $[D_j]_0=100$. (b) $W_\infty$ as a function of the initial door concentration $[D_j]_0$ (numerical only). Parameters: $[P]_0=1$ and $k_{\mathrm{st}}=k_{\mathrm{sw}}=k_1=k_2=1$.}
  \label{fig:SteadyStates}
\end{figure}

The reaction outcome for the mechanism with two competing strategies is shown in Figure~\ref{fig:SteadyStates}a. Here the rate constant \(k_{\mathrm{sw}}\) is kept constant at 1, while \(k_{\rm st}\) is varied between \(10^{-3}\) and \(10^{3}\). We find that the winning species concentration decreases monotonically from \(2/3\) to \(1/3\) as \(k_{\rm st}\) increases. A winning species concentration of 0.5 is found for \(k_{\mathrm{st}} \approx 1\) if the initial door concentrations are high, as shown in Figure~2.

Since each final species requires three D's for each P, low initial D concentrations result in some players not being able to complete the game. To evaluate the effect of this limiting reactant scenario on our model, we systematically varied the initial concentration of each door species between 0 and 6, corresponding to total concentrations of 0 to 18. The results are shown in Figure~\ref{fig:SteadyStates}b. We find that a pseudo-first-order description becomes appropriate for \([D_{j}]_0  \gtrsim 5\) to 6.

\section{Analytical Solutions}
\label{sec:analytical}

\noindent Since we set \(g=1\) (prize behind door~1), the always-switch cascade starting from \(I_1\) splits into two losing pathways at the host’s reveal, whereas the two cascades beginning with a goat pick (\(j=2,3\)) remain unbranched (Figure~\ref{fig:Schematic}).  Under pseudo-first-order conditions \([D_j]_0=D_0\gg P_0\) and with no stay step (\(k_{\mathrm{st}}=0\)), these winning cascades obey the same rate law.  We arbitrarily select the \(I_3\) cascade, which accounts for exactly half of the win production, and define \(\displaystyle \mathbf c(t) = \bigl([P],\,[I_3],\,[I_{3,2}],\,[I_{3,2,1}]\bigr)^T\). This concentration vector obeys

\begin{equation}
\label{eq:matrixequation}
\frac{d\mathbf c}{dt}
= D_0
\begin{pmatrix}
-3k_1 & 0         & 0                   & 0\\[6pt]
 k_1 & -k_2       & 0                   & 0\\[6pt]
 0    & k_2       & -k_{\mathrm{sw}}    & 0\\[6pt]
 0    & 0         & k_{\mathrm{sw}}     & 0
\end{pmatrix}
\mathbf c(t).
\end{equation}

\noindent With the initial condition \(\mathbf{c}(0)=[P_0,0,0,0]^T\), a Laplace-transform \cite{zhou2007} or matrix-exponential calculation yields a closed-form solution for the time evolution of \([I_{3,2,1}]\). Adding the second winning cascade, the resulting win is twice that value and hence

\begin{equation}
\label{eq:Wt}
\begin{split}
W(t)
&=2~P_0 \bigg[
\frac{k_1 k_{\mathrm{sw}} e^{-D_0 k_2 t}}
     {(k_2 - k_{\mathrm{sw}})\,(3k_1 - k_2)}
\;-\;
\frac{k_1 k_2 e^{-D_0 k_{\mathrm{sw}} t}}
     {(k_2 - k_{\mathrm{sw}})\,(3k_1 - k_{\mathrm{sw}})}\\
&\quad 
\;-\;\frac{k_2 k_{\mathrm{sw}} e^{-3D_0 k_1 t}}
     {(3k_1 - k_2)\,(9k_1 - 3k_{\mathrm{sw}})}
\;+\;
\frac{1}{3}\,\bigg].
\end{split}
\end{equation}

\noindent This function is zero at \(t=0\) and converges to the equilibrium concentration \(\tfrac{2}{3}P_0\).

Notice that the denominators in eq.~\ref{eq:Wt} can become zero, producing indeterminate forms that are resolvable using l’Hôpital’s rule. The closed-form result for \(k_2 = k_{\rm sw}\) is given in Appendix~\ref{app:degenerate} and superposed in Figure~\ref{fig:Kinetics}a, where it matches the numerical result. Similarly, the always-stay winner concentration follows the same form as the switch-loss curve (eq.~\ref{eq:Wt} with \(k_{\rm sw}\to k_{\rm st}\)) but converges to \(\tfrac{1}{3}P_0\) (Figure~\ref{fig:Kinetics}b).  

Lastly, we can analyze the mixed-strategy case. In the long‑time limit, a branch that began with the correct initial guess (\(j=g\), probability \(1/3\)) wins by staying with probability \(k_{\mathrm{st}}/(k_{\mathrm{sw}}+k_{\mathrm{st}})\), whereas a branch that began with a goat (\(j\neq g\), probability \(2/3\)) wins by switching with probability \(k_{\mathrm{sw}}/(k_{\mathrm{sw}}+k_{\mathrm{st}})\).  
Hence the total equilibrium win concentration is
\begin{equation}
\label{eq:Winf}
W_\infty
= P_{0}\,\frac{k_{\mathrm{st}} + 2\,k_{\mathrm{sw}}}{3\,(k_{\mathrm{sw}} + k_{\mathrm{st}})}.
\end{equation}
This expression is overlaid on the numerical data in Figure~\ref{fig:SteadyStates}a, showing excellent agreement. Notice that 
$W_\infty$ is independent of $k_1$ and $k_2$ which only set the time scale required for establishing the final state.

\section{Polymer Interpretation}
In the context of the proposed reaction mechanisms, the player species can be interpreted as growing linear polymers, for which \(P\) serves as a molecular scaffold.  Upon selecting door \(j\), the intermediate \(I_j\) corresponds to the conjugate \(P\!-\!D_j\).  After the host reveals door \(k\), the structure extends to \(P\!-\!D_j\!-\!D_k\), forming \(I_{j,k}\).  The final decision adds a third unit, resulting in the fully extended product \(I_{j,k,\ell}\), or \(P\!-\!D_j\!-\!D_k\!-\!D_\ell\).

This oligomer chemically encodes the entire decision history, with each monomer uniquely representing a specific step: the initial choice, the host’s reveal, and the final decision. Accordingly, each of the four final products carries \(\log_2(4) = 2\) bits of information. Beyond the Monty Hall puzzle, this approach generalizes to any finite decision tree, where each branching event adds a distinct monomer. Because the reaction steps do not involve nonlinearities or feedback loops, the resulting kinetics form simple cascades of linear rate laws and yield analytically solvable outcomes that directly encode the decision probabilities.

Potential implementations of these decision polymers demand orthogonality between reaction steps and the independent tunability of rate constants. DNA strand-displacement reactions can--within limits--offer these capabilities, as discussed in the following section.

\section{Proposed Experimental Approach}

DNA strand displacement using synthetic oligonucleotides \cite{soloveichik2010,qian2011} provides a modular, sequence-programmable route to implement our Monty Hall network. Each species is realized as a tailored DNA ``gate,'' typically a duplex or hairpin exposing a short single-stranded toehold that nucleates branch migration, driving effectively irreversible strand-exchange steps. This strategy, and its extensions, have enabled the engineering of intricate reaction mechanisms including synthetic oscillators, bistable switches, and autocatalytic cascades \cite{lv2021}.

Since the Monty Hall network comprises three concurrent cascades (pick, reveal, and decision), circuit fidelity critically depends on maintaining truly orthogonal toeholds. Non-orthogonal sequences risk unintended cross-interactions, compromising cascade specificity. Additionally, effective rate constants must be precisely tuned, and active management of reaction by-products (e.g., via leak-suppressing blockers or enzymatic cleanup) is essential to prevent accumulation of spent strands. Fluorogenic reporters or unique sequence barcodes can facilitate real-time monitoring of product formation.

Under these conditions, the length of each toehold (typically 1-8 nucleotides) directly tunes the effective rate constant for branch-migration initiation, allowing fine control over each step \cite{zhang2009}. Specifically, the cascade proceeds by first forming intermediate \(I_j\) when the door strand \(D_j\) binds to scaffold strand \(P\). Subsequently, a reveal strand \(D_k\) invades via its distinct toehold, yielding intermediate \(I_{j,k}\). Finally, the decision strand \(D_\ell\) completes the cascade, producing \(I_{j,k,\ell}\), whose sequence explicitly encodes the entire decision pathway. With careful sequence design, kinetic optimization, and efficient waste management, DNA strand displacement could provide a robust, programmable testbed for our chemically encoded Monty Hall decision process.

\section{Discussion and Conclusion}

We have shown that the Monty Hall problem and its characteristic always-stay and always-switch strategies can be encoded in simple chemical reaction mechanisms. The resulting mass-action kinetics reproduce the expected probabilities of 1/3 for staying and 2/3 for switching as equilibrium concentrations of specific chemical species.

Our approach complements recent treatments of the Monty Hall problem using quantum-mechanical frameworks that exploit probabilistic amplitudes and entanglement \cite{flitney2002,kurzyk2016}, and DNA-based implementations employing massively parallel sequencing \cite{mamet2019}. By developing a deterministic, reaction-kinetic formulation, our study bridges classical probability theory, chemical computation, and decision-making, offering a new platform for encoding logical strategies at the molecular scale.

Our findings highlight a fundamental form of passive chemical decision-making under simple linear rate laws:  the partitioning of flux among competing pathways without active sensing or feedback. Such mechanisms might have played a role in prebiotic chemistry and early life because sensory machinery was absent yet strategic resource allocation necessary \cite{muchowska2020}. Similar minimalist decision-making strategies are also common in engineering. For example, high-energy physics detectors use rapid analog hardware to apply simple logic and probabilistic rules to collision data incoming at rates exceeding one billion events per second \cite{cernTrigger}. These trigger systems efficiently preselect only the most promising events for deeper analysis.

A limitation of our proof-of-principle is the combinatorial growth in species and reactions, as even the always-switch scenario alone involves 15 distinct species and 11 core reactions. Scaling to more complex decision problems will therefore require strategies to reduce network overhead, such as modular or hierarchical designs, catalytic reuse of intermediates, or template-mediated architectures.

In conclusion, our reaction-kinetic model of the Monty Hall problem offers a minimalist blueprint for molecular‐scale decision-making, embedding strategy in kinetic asymmetries so that chemical systems select advantageous pathways without active sensing or computation.  This parallels simple engineering triggers and suggests that early biological systems likewise exploited kinetic biases for adaptation before complex regulation evolved. Further studies of these passive frameworks may uncover hard-wired decision architectures within biochemical networks.

\section{Methods}

All simulations and analyses were performed in MATLAB R2024b. The kinetic rate equations were integrated with the built-in \texttt{ode45} solver (based on the Dormand-Prince explicit Runge-Kutta method) \cite{shampine1997}. The corresponding MATLAB scripts used to generate Figures~\ref{fig:Kinetics} and~\ref{fig:SteadyStates}, as well as a symbolic-math script that derives and verifies the closed-form solutions of eqs.~\ref{eq:Wt} and~\ref{eq:twoequal}, are available at \url{https://github.com/osteinbock/MontyHall}.

\newpage

\appendix

\section{Appendix: Chemical Species}
\label{app:species-list}
Our model involves the following 19 species.
\vspace{1\baselineskip}

\noindent
\begin{tabular}{@{}l@{\quad}l@{}}
Player:                   & $P$ \\
Doors:                    & $D_1, D_2, D_3$ \\
Stage 1 (pick):           & $I_1, I_2, I_3$ \\
Stage 2 (reveal):         & $I_{1,2}, I_{1,3}, I_{2,3}, I_{3,2}$ \\
Final products (switch):  & $I_{1,2,3}, I_{1,3,2}, I_{2,3,1}, I_{3,2,1}$ \\
Final products (stay):    & $I_{1,2,1}, I_{1,3,1}, I_{2,3,2}, I_{3,2,3}$
\end{tabular}

\section{Appendix: Rate Laws}
\label{app:rate-laws}

Our model is governed by the following 19 rate laws. Note that in each sum over reveal‐indices (e.g.\ $k\neq j$), we implicitly exclude $k=g$, since no species $I_{j,k}$ (or $I_{j,k,\ell}$) is defined for $k=g$ (the host never reveals the prize door).

\begin{flalign}
&\frac{d[P]}{dt}
  = -k_1\sum_{j=1}^3 [P][D_{j}],                                                   &&\\
&\frac{d[D_{j}]}{dt}
  = -k_1[P][D_{j}]
    - k_2\sum_{\substack{i=1\\ i\neq j}}^3 [I_{i}][D_{j}]
    - k_{\mathrm{sw}}\sum_{\substack{p,q=1\\ p\neq q}}^3 [I_{p,q}][D_{j}]
    - k_{\mathrm{st}}\sum_{\substack{s=1\\ s\neq j}}^3 [I_{j,s}][D_{j}],           &&\\
&\frac{d[I_{j}]}{dt}
  = k_1[P][D_{j}]
    - k_2\sum_{\substack{k=1\\ k\neq j}}^3 [I_{j}][D_{k}],                           &&\\
&\frac{d[I_{j,k}]}{dt}
  = k_2[I_{j}][D_{k}]
    - k_{\mathrm{st}}[I_{j,k}][D_{j}]
    - k_{\mathrm{sw}}\sum_{\substack{\ell=1\\ \ell\neq j,k}}^3 [I_{j,k}][D_{\ell}],  &&\\
&\frac{d[I_{j,k,\ell}]}{dt}
  = k_{\mathrm{sw}}[I_{j,k}][D_{\ell}],  \qquad \text{for } \ell \neq j,\ k          \\[\baselineskip]
&\frac{d[I_{j,k,j}]}{dt}
  = k_{\mathrm{st}}[I_{j,k}][D_{j}], \qquad \text{with } j \neq k.                   &&
\end{flalign}

\section{Appendix: Identical Rate Constants}
\label{app:degenerate}

The expression for \(W(t)\) in eq.~\eqref{eq:Wt} contains removable singularities whenever two of the three pseudo-first-order rate constants coincide.  In each case, one can apply l’Hôpital’s rule. For \(k_2 = k_{\mathrm{sw}} \neq 3k_1\) (\(k_{\rm st}=0\)), the cumulative win concentration is 

\begin{equation}
\label{eq:twoequal}
\begin{aligned}
W_{k_2=k_{\mathrm{sw}}}(t) &= 2 P_0 \bigg[\frac{k_{\mathrm{sw}}^{2}\,e^{-3D_{0}k_{1}t}\,e^{-D_{0}k_{\mathrm{sw}}t}\,\bigl(e^{3D_{0}k_{1}t}-e^{D_{0}k_{\mathrm{sw}}t}\bigr)}
     {3\,(3k_{1}-k_{\mathrm{sw}})^{2}}\\[6pt]
       &\quad - \frac{D_{0}\,k_{\mathrm{sw}}^{2}\,t\,e^{-D_{0}k_{\mathrm{sw}}t}}
         {3\,(3k_{1}-k_{\mathrm{sw}})}
         - \frac{e^{-D_{0}k_{\mathrm{sw}}t}\,\bigl(D_{0}k_{\mathrm{sw}}t - e^{D_{0}k_{\mathrm{sw}}t} + 1\bigr)}
         {3}\,\bigg].
\end{aligned}
\end{equation}


\newpage


\begin{thebibliography}{7}

\bibitem{tyson2014}
J.~J.~Tyson and B.~Novak, \emph{Interface Focus}, 2014, \textbf{4}, 20130070.
 DOI: 10.1098/rsfs.2013.0070.

\bibitem{costa2012}
P.~Costa, F.~V.~M.~Almeida and J.~T.~Connelly, \emph{Int.\ J.\ Biochem.\ Cell Biol.}, 2012, \textbf{44}, 2233-2237, DOI: 10.1016/j.biocel.2012.09.003.

\bibitem{kay2008}
R.~Kay, P.~Langridge, D.~Traynor \emph{et al.}, \emph{Nat.\ Rev.\ Mol.\ Cell Biol.}, 2008, \textbf{9}, 455-463, DOI:10.1038/nrm2419.

\bibitem{friedl2003}
P.~Friedl and K.~Wolf, \emph{Nat.\ Rev.\ Cancer}, 2003, \textbf{3}, 362-374, DOI: 10.1038/nrc1075.

\bibitem{delvecchio2016}
D.\ Del Vecchio, A.\ J.\ Dy and Y.\ Qian, \emph{J.\ R.\ Soc.\ Interface}, 2016, \textbf{13}, 20160380. DOI: 10.1098/rsif.2016.0380.

\bibitem{hjelmfelt1991}
A.\ Hjelmfelt, E.\ D.\ Weinberger and J.\ Ross, \emph{Proc.\ Natl.\ Acad.\ Sci.\ U.\ S.\ A.}, 1991, \textbf{88}, 10983-10987. DOI: 10.1073/pnas.88.24.10983

\bibitem{hjelmfelt1992a}
A.\ Hjelmfelt and J.\ Ross, \emph{Proc.\ Natl.\ Acad.\ Sci.\ U.\ S.\ A.}, 1992, \textbf{89}, 388-391. DOI: 10.1073/pnas.89.1.388.

\bibitem{hjelmfelt1992b}
A.\ Hjelmfelt, E.\ D.\ Weinberger and J.\ Ross, \emph{Proc.\ Natl.\ Acad.\ Sci.\ U.\ S.\ A.}, 1992, \textbf{89}, 383-387. DOI: 10.1073/pnas.89.1.383.

\bibitem{zhu2024}
W.~Zhu, P.~Knoll and O.~Steinbock, \emph{J. Phys. Chem. Lett.}, 2024, \textbf{15}, 5476--5487, DOI: 10.1021/acs.jpclett.4c01031.

\bibitem{steinbock1995}
O.~Steinbock, Á.~Tóth and K.~Showalter, \emph{Science}, 1995, \textbf{267}, 868-871, DOI: 10.1126/science.267.5199.868.

\bibitem{beta2017}
C.\ Beta and K.\ Kruse, \emph{Annu.\ Rev.\ Condens.\ Matter Phys.}, 2017, \textbf{8}, 239-264. DOI: 10.1146/annurev-conmatphys-031016-025210

\bibitem{selvin1975a}
S.~Selvin, \emph{The American Statistician}, 1975, \textbf{29}, 67-71, DOI: 10.1080/00031305.1975.10479121.

\bibitem{whitmeyer2017}
M.~Whitmeyer, \emph{Games}, 2017, \textbf{8}(3), 31, DOI: 10.3390/g8030031.

\bibitem{rosenhouse2019}
J.~Rosenhouse, \emph{The Monty Hall Problem: The Remarkable Story of Math’s Most Contentious Brain Teaser}, Oxford University Press, Oxford, 2019.

\bibitem{krauss2003}
S.~Krauss and X.~T.~Wang, \emph{J. Exp. Psychol. Gen.}, 2003, \textbf{132}, 3-22, DOI: 10.1037/0096-3445.132.1.3.

\bibitem{wikipedia}
Wikipedia contributors, “Monty Hall problem,” \emph{Wikipedia, The Free Encyclopedia}, \url{https://en.wikipedia.org/wiki/Monty_Hall_problem}, accessed April 27, 2025.

\bibitem{zhou2007}
Y.\ Zhou and X.\ Zhuang, \emph{J.\ Phys.\ Chem.\ B}, 2007, \textbf{111}, 13600-13610. DOI: 10.1021/jp073708+.

% \bibitem{hill2009}
% T.~P.~Hill, \emph{American Scientist}, 2009, \textbf{97}, 126-133, DOI: 10.1511/2009.77.126.

\bibitem{soloveichik2010}
D.~Soloveichik, G.~Seelig and E.~Winfree, \emph{Proc.\ Natl.\ Acad.\ Sci.\ U.\ S.\ A.}, 2010, \textbf{107}, 5393-5398, DOI: 10.1073/pnas.0909380107.

\bibitem{qian2011}
L.~Qian, E.~Winfree and J.~Bruck, \emph{Nature}, 2011, \textbf{475}, 368-372, DOI: 10.1038/nature10262.

\bibitem{lv2021}
H.\ Lv, Q.\ Li, J.\ Shi, C.\ Fan and F.\ Wang, \emph{ChemPhysChem}, 2021, \textbf{22}, 1151-1166. DOI: 10.1002/cphc.202100140.

\bibitem{zhang2009}
D.\ Y.\ Zhang and E.\ Winfree, \emph{J.\ Am.\ Chem.\ Soc.}, 2009, \textbf{131}, 17303-17314. DOI: 10.1021/ja906987s.

\bibitem{flitney2002}
A.~P.~Flitney and D.~Abbott, \emph{Phys. Rev. A}, 2002, \textbf{65}, 062318, DOI: 10.1103/PhysRevA.65.062318.

\bibitem{kurzyk2016}
D.~Kurzyk and A.~Glos, \emph{Quantum Inf. Process.}, 2016, \textbf{15}, 4927-4937, DOI: 10.1007/s11128-016-1431-8.

\bibitem{mamet2019}
N.\ Mamet, G.\ Harari, A.\ Zamir and I.\ Bachelet, \emph{Comput.\ Biol.\ Chem.}, 2019, \textbf{83}, 107122. DOI: 10.1016/j.compbiolchem.2019.107122.

\bibitem{muchowska2020}
K.\ B.\ Muchowska, S.\ J.\ Varma and J.\ Moran, \emph{Chem.\ Rev.}, 2020, \textbf{120}(15), 7708-7744. DOI: 10.1021/acs.chemrev.0c00191.

\bibitem{cernTrigger}
CERN, ``ATLAS Trigger and Data Acquisition,'' \emph{ATLAS Public Website}, 2024, \url{https://atlas-public.web.cern.ch/discover/detector/trigger-daq}, accessed April 27, 2025.

\bibitem{shampine1997}
L.\ F.\ Shampine and M.\ W.\ Reichelt, \emph{SIAM J.\ Sci.\ Comput.}, 1997, \textbf{18}, 1-22. DOI: 10.1137/S1064827594276424.

\end{thebibliography}
\end{document}